\documentclass[%
 reprint,
superscriptaddress,
 amsmath,amssymb,
 aps,pra
]{revtex4-2}

\usepackage{graphicx}
\usepackage{dcolumn}
\usepackage{bm}
\usepackage[colorlinks=true,linkcolor=blue,urlcolor=blue,citecolor=blue]{hyperref}

\usepackage{subfigure}
\usepackage{xcolor}
\usepackage{amsmath}
\usepackage{physics}
\usepackage{float}

\usepackage[english]{babel}
\begin{document}

\preprint{APS/123-QED}
\title{Inhomogeneous spin momentum induced orbital motion of birefringent particles in tight focusing of vector beams in optical tweezers}

 \author{Ram Nandan Kumar}
 \email{ramnandan899@gmail.com}
 \affiliation{Department of Physical Sciences, Indian Institute of Science Education and Research Kolkata, Mohanpur-741246, West Bengal, India}

 \author{Sauvik Roy}
 \affiliation{Department of Physical Sciences, Indian Institute of Science Education and Research Kolkata, Mohanpur-741246, West Bengal, India}

 \author{Anand Dev Ranjan}
 \affiliation{Department of Physical Sciences, Indian Institute of Science Education and Research Kolkata, Mohanpur-741246, West Bengal, India}

\author{Subhasish Dutta Gupta}
\affiliation{Department of Physical Sciences, Indian Institute of Science Education and Research Kolkata, Mohanpur-741246, West Bengal, India}
\affiliation{Department of Physics, Indian Institute of Technology, Jodhpur 342030, India}
\affiliation{Tata Institute of Fundamental Research Hyderabad, India}

\author{Nirmalya Ghosh}
\affiliation{Department of Physical Sciences, Indian Institute of Science Education and Research Kolkata, Mohanpur-741246, West Bengal, India}

\author{Ayan Banerjee}
 \email{ayan@iiserkol.ac.in}
 \affiliation{Department of Physical Sciences, Indian Institute of Science Education and Research Kolkata, Mohanpur-741246, West Bengal, India}

\date{\today}

\begin{abstract}

Spin orbit interaction (SOI) due to tight focusing of light in optical tweezers has led to exciting and exotic avenues towards inducing rotation in microscopic particles. However, instances where the back action of the particles influences and modifies SOI effects so as to induce rotational motion are rarely known.  Here, we tightly focus a vector beam having radial/azimuthal polarization carrying no intrinsic angular momentum, into a refractive index stratified medium, and observe orbital rotation of birefringent particles around the beam propagation axis. In order to validate our experimental findings, we perform numerical simulations of the underlying equations. Our simulations reveal that the interaction of light with a birefringent particle gives rise to inhomogeneous spin currents near the focus, resulting in a finite spin momentum. This spin momentum combines with the canonical momentum to finally generate an origin-dependent orbital angular momentum which is manifested in the rotation of the birefringent particles around the beam axis. Our study describes a unique modulation of the SOI of light due to interaction with anisotropic particles that can be used to identify new avenues for exotic and complex particle manipulation in optical tweezers.

\end{abstract}


\maketitle

\section{Introduction}

Optical tweezers have facilitated both spin (about an axis fixed to the particle) \cite{friese1998optical,stilgoe2022controlled} and orbital rotation (about the axis of the trapping laser beam) of trapped single or multiple particles  \cite{curtis2003structure}. While the spinning motion around the particle axes is due to exchange of the longitudinal or transverse spin angular momentum ($S_z$ or ${S}_{\perp}$) of the beam with birefringent particles \cite{friese1998optical,stilgoe2022controlled,kumar2023probing}, the orbital motion is principally caused by tight focusing of Laguerre-Gauss (LG) or Bessel-Gauss beams carrying intrinsic OAM \cite{garces2003observation, zhao2007spin,zhan2009cylindrical}. In addition, over the past couple of decades, researchers have revealed various exotic effects of the spin-orbit interaction (SOI) of light in scattering, imaging, and beam propagation through anisotropic media, as well as in the tight focusing of light \cite{ray2017polarization,forbes2019structured,singh2018transverse,fu2019spin,nayak2023spin}. Notably, manifestations of SOI, including orbital motion, have been observed by tightly focusing a spin-polarized (with positive or negative helicity) Gaussian beam through an RI-stratified medium to generate significant transverse momentum. This momentum even bears the signatures of the elusive Belinfante spin \cite{pal2020direct}. 

On another note, particles trapped using tweezers scatter a significant amount of light in the longitudinal direction (both backward and forward) - which is used to characterize the trap stiffness rather efficaciously \cite{neuman2004optical,pal2012measurement}. The scattered light also leads to optical binding effects, which enables the trapping of multiple particles and generates both attractive and repulsive inter-particle forces that have been well-studied \cite{dholakia2010colloquium}. However, to the the best of our knowledge, the effects of light-matter interactions in modulating the SOI of light so as to produce dynamics in trapped particles has not been well-researched. This is of particular consequence in the case of birefringent particles, which have the ability to modify the polarization of incident light. Thus, any inhomogeneous polarization generated by the SOI of light would be further altered in the presence of birefringent particles, giving rise to intriguing effects. Here, the choice of input polarization of the light would play a crucial role too, with inhomogeneously polarized (radial/azimuthal polarization) light likely to accentuate the observed effects even further.

This is indeed our focus of research in this paper. Thus, we employ structured vector beams as the input into an optical tweezers system. Since tight focusing inevitably generates a significant longitudinal field component, our expectation is that this would be likely to result in unexpected effects in the case of vector beams \cite{friese1998optical,yang2021optical,kumar2022probing, forbes2021structured}. The vector beams we employ are first-order radially and azimuthally polarized, and are characterized by an intensity singularity on the beam axis and unique properties of cylindrical symmetry in their polarization \cite{machavariani2007efficient, andrews2012angular, zhan2009cylindrical, huang2011vector}. Such properties facilitate low intensity loss during propagation, even under non-paraxial conditions \cite{rodriguez2010optical, dorn2003sharper}. Understandably, the consequences of tight focusing such beams on birefringent microparticles deserve attention \cite{youngworth2000focusing,kumar2023probing}. Additionally, in our case, the tightly focused beams also propagate through a stratified refractive index (RI) medium before entering the trapping region inside our sample chamber \cite{roy2013controlled}. The influence of the stratified medium is actually crucial in determining the interaction of the light beam with the particles and influencing their dynamics, as we had observed earlier in Refs.~\cite{roy2013controlled,pal2020direct}.  In the present case, for input radially or azimuthally polarized light, the canonical momentum and the spin momentum (SM) together generate a substantial origin-dependent (extrinsic) OAM of light, which we exploit to rotate trapped birefringent particles in the transverse plane around the beam axis  \cite{o2002intrinsic,novotny2001longitudinal}. The effect of the spin momentum is apparent from the orbital motion readily induced in birefringent particles, while polystyrene microspheres (non-birefringent) of the same diameter appear unaffected. The spin momentum is generated due to the fact that the interaction of the inhomogeneous input polarization at the trap focal plane with birefringent particles - having a high value of optical retardance - lead to the generation of spin currents (regions of elliptical polarization) around the beam axis. We explain all observed experimental effects by carrying out rigorous numerical simulations of the equations governing our system for different RI values of the stratified medium that the tightly focused beam encounters. Our work provides an experimentally viable strategy for engineering optical traps with controlled and specific orbital motion of trapped particles by SOI effects generated by tight focusing alone, without the need for complex algorithms involving adaptive optics. In what follows, we first describe our experiments and observations.

\begin{center}
\begin{figure}[!h]
\includegraphics[width=0.5\textwidth]{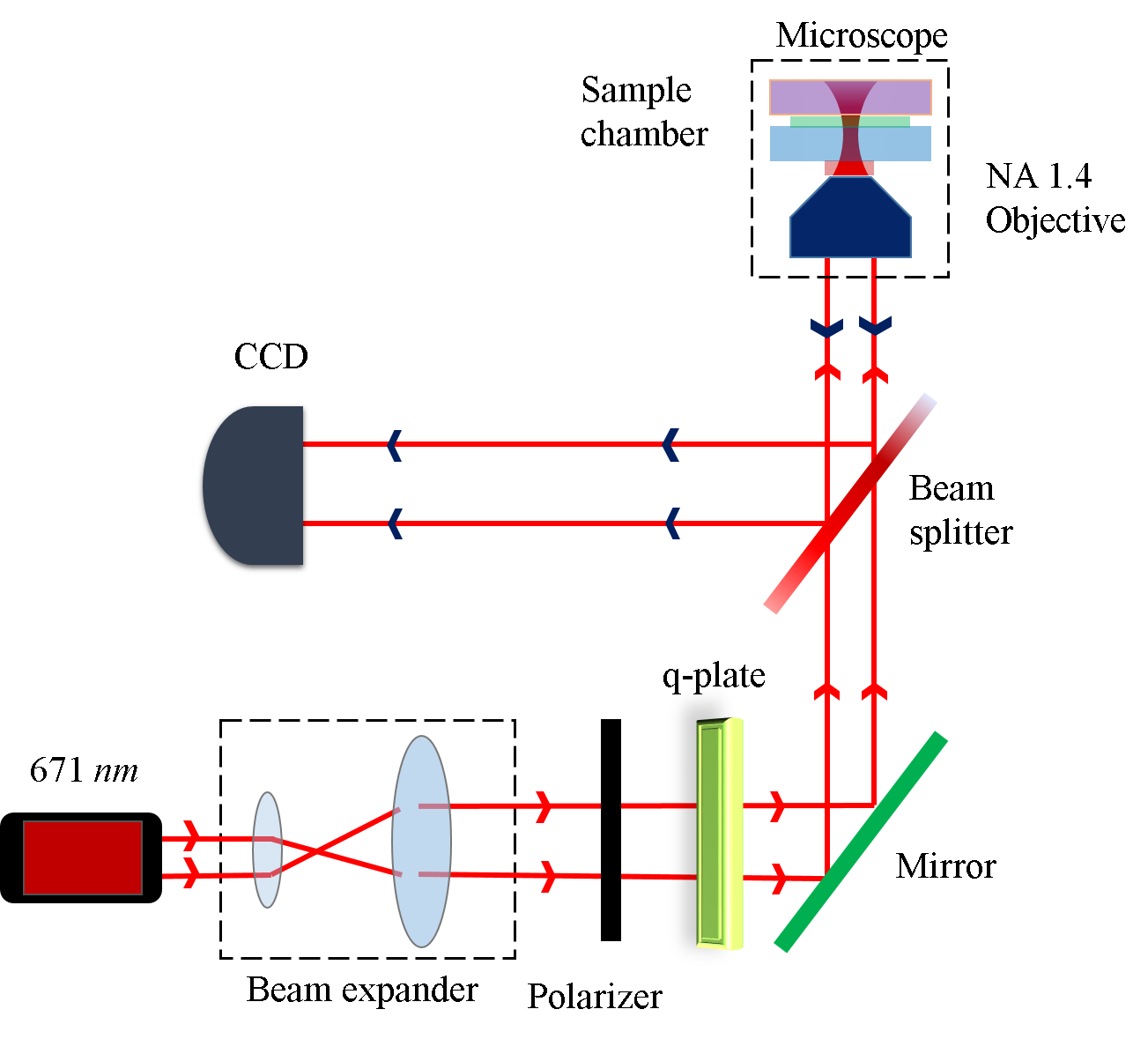}
\caption{Schematic representation of the experimental setup designed to couple radially or azimuthally polarized light into our optical tweezers system for conducting experiments on birefringent RM257 particles.}
\label{exp_schematic}
\end{figure}
\end{center}

\section{EXPERIMENTAL RESULTS}

The schematic and details of our optical tweezers setup are shown in Fig.~\ref{exp_schematic}.  Thus, we use a conventional optical tweezers configuration consisting of an inverted microscope (Carl Zeiss Axioert.A1) with an oil-immersion 100X objective (Zeiss, NA 1.4) and a solid-state laser (Lasever, 671 $nm$, 350 $mW$) coupled to the back port of the microscope. We use a vortex half-wave retarder (q-plate) of zero-order for generating structured vector beams (i.e. first-order radially and azimuthally polarized vector beams). We fix the fast axis orientation of the vortex plate ($q=\frac{1}{2}$) in such a way that it converts linear $x$- polarized and $y$-polarized light into azimuthally and radially polarized light, respectively \cite{marrucci2006optical,yan2015q,kumar2022manipulation}. For the probe particles, we first fabricate the  RM257 vaterite, which is optically anisotropic and birefringent. The dielectric constants ${\epsilon}_{||}$ and ${\epsilon}_{\perp}$ are related to the ordinary and extraordinary refractive indices ${n}_{||}$ and ${n}_{\perp}$ of the birefringent particle through the relationships ${n}^2_{||}= {\epsilon}_{||}$ and ${n}^2_{\perp}= {\epsilon}_{\perp}$. In the case of RM257 birefringent particles, there is an approximate difference of 0.13 between the ordinary and extraordinary refractive indices \cite{sandomirski2004highly}. Subsequently, we quantify the linear retardance ($\delta$) of the RM257 particles by analyzing the Mueller matrix. This measured value provides valuable insights into the OAM transfer dynamics, allowing us to probe the influence of OAM in the experimental setup comprehensively. We then couple the radially (azimuthally) polarized vector beam into the microscope so that it is tightly focused into the stratified medium. The cover slip and the glass slide sandwiched together comprise the sample chamber into which we add approximately $20 ~\mu l$ of the aqueous dispersion of the RM257 particles. The mean size of these particles is 2-3 ~$\mu m$ with a standard deviation of 20\%. We collect the forward-transmitted light from the microscope lamp, as well as back-reflected light from the RM257 particles, to characterize the orbital motion. The OAM transfer to particles trapped at the trap center is optimized by varying the $z$-focus of the microscope objective.

\begin{center}
\begin{figure}[!h]
\includegraphics[width=0.5\textwidth]{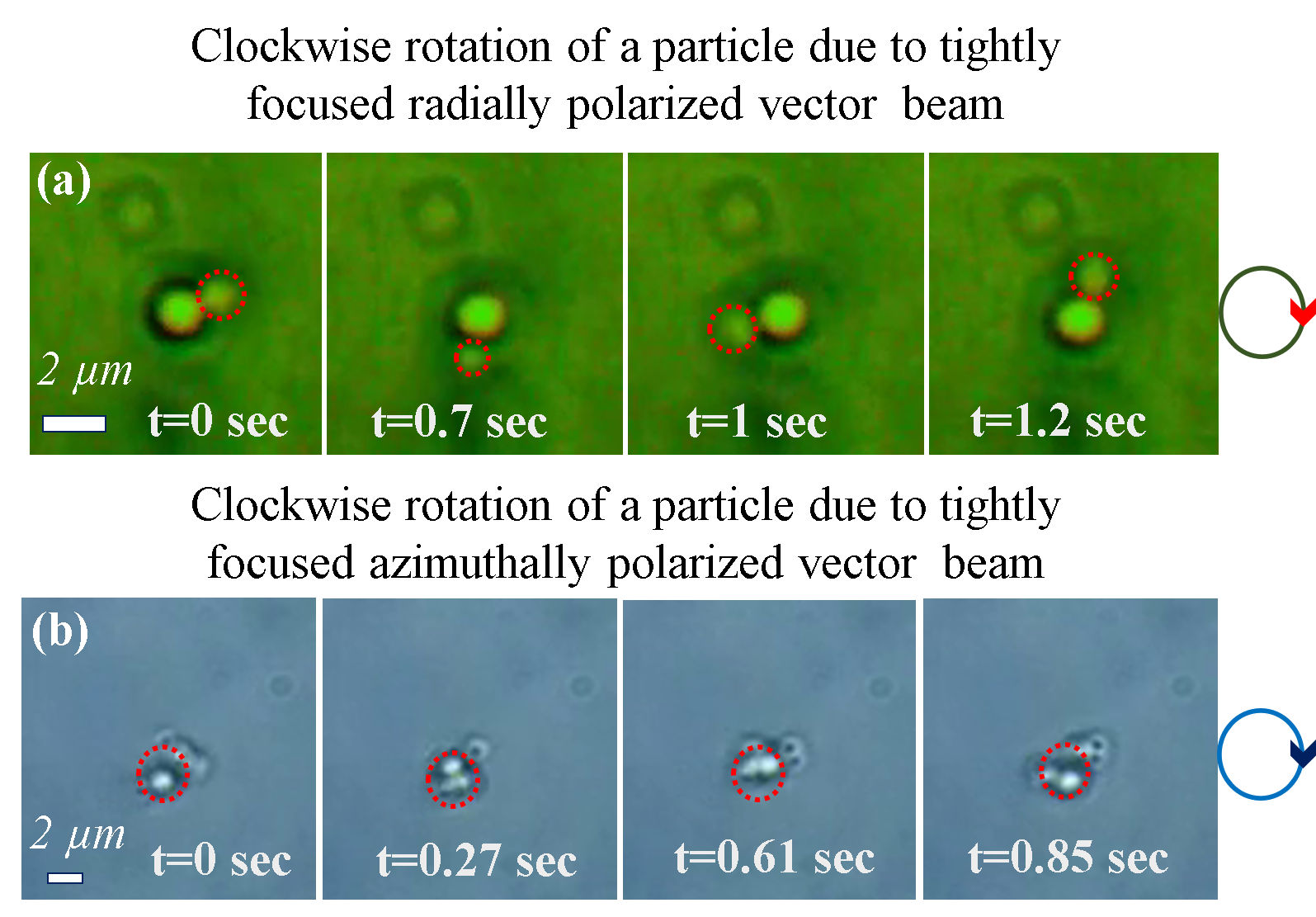}
\caption{ Time-lapsed frames from video recordings depict the rotation of particles induced by a tightly focused radially (azimuthally) polarized vector beam. (a) The trajectory of an RM257 birefringent particle rotating around another trapped particle at the center of the beam is marked by red circles, even though the two particles are at slightly different axial depths. (b) The red circle show the orbit of rotation of similar particles at 2$\mu$m away from the focus of the azimuthally polarized vector beam. It's noteworthy that there is zero intensity ($E_{z}=0$) at the center of the beam. As a result, particles are orbiting solely in an annular ring of the intensity. }
\label{exp_OAM}
\end{figure}
\end{center}

The experimental results are shown in Figs.~\ref{exp_OAM}(a) and (b). We probe the effect of total OAM by trapping birefringent RM257 particles using radially and azimuthally polarized vector beams. In Fig.~\ref{exp_OAM}(a), we observe that a single large particle (size around 2 $\mu$m) is trapped at the beam center, while another smaller particle (size around 1.5 $\mu$m) is rotating in the clockwise direction in the annular ring in time lapsed images. In the case of azimuthally polarized light, smaller RM257 particles are not trapped at the center of the beam because the longitudinal component of the electric field is zero, so we observe single particles, as well as clusters, rotating clockwise around the beam center, as we show in the time lapsed images in Fig.~\ref{exp_OAM}(b). Thus, these experiments clearly demonstrate that tight focusing of radially and azimuthally polarized light generates OAM that helps in rotating the particles about the beam propagation axis, even though the beam does not contain any spin angular momentum (SAM) and intrinsic OAM.  We also observe that the frequency of rotation increases as we increase the power. This is expected as the magnitude of both the electric and magnetic fields will increase on increasing the input beam intensity. We now attempt to explain our experimental observations by numerical simulations - which we perform using a theoretical formalism that we describe in the next section.

\section{THEORETICAL CALCULATIONS}
Tight focusing due to objective lenses with a high NA  generates a non-paraxial condition \cite{kumar2022probing,Novotny2012,kumar2021study}. For the determination of electric and magnetic fields of radially and azimuthally polarized vector beams under such non-paraxial conditions, we use the angular spectrum method or vector diffraction theory provided by Richards and Wolf \cite{zhao2007spin,sato2009hollow,richards1959electromagnetic}.
The general expression of Cartesian components of the electric-field ($E_{x}$, $E_{y}$, and $E_{z}$) of a focused radially polarized vector beam in the focal plane  is given as
\onecolumngrid

\begin{multline}
\begin{split}
\begin{aligned}
\left[\begin{array}{l}
E_{x}^{o} \\
E_{y}^{o} \\
E_{z}^{o}
\end{array}\right]_{R}=A i^{m+1} \exp (i m \phi) \int_{0}^{\theta_{\max }} f_{\omega}(\theta) \cos ^{3 / 2} \theta \sin ^{2}  \theta \exp (i k z \cos \theta)  \\ \left[\begin{array}{l}
-i\left(J_{m+1}(\beta) - J_{m-1}(\beta)\right) \cos \phi+\left(J_{m+1}(\beta)+J_{m-1}(\beta)\right) \sin \phi \\
-i\left(J_{m+1}(\beta) - J_{m-1}(\beta)\right) \sin \phi-\left(J_{m+1}(\beta)+J_{m-1}(\beta)\right) \cos \phi \\
~~~~~~~~~~~~~~~~~~~2\tan \theta J_{m}(\beta)
\end{array}\right] d \theta
\label{rad}
\end{aligned}
\end{split}
\end{multline}

Similarly, the general form of the Cartesian components of of the electric field of the azimuthally polarized vector beam can be expressed as

\begin{multline}
    \begin{aligned}
\left[\begin{array}{l}
E_{x}^{o} \\
E_{y}^{o} \\
E_{z}^{o}
\end{array}\right]_{A}=A i^{m+1} \exp (i m \phi) \int_{0}^{\theta_{\max }} f_{\omega}(\theta) \cos ^{1/ 2} \theta \sin ^{2} \theta \exp (i k z \cos \theta) \\ 
\left[\begin{array}{l}
\left(J_{m+1}(\beta)+J_{m-1}(\beta)\right) \cos \phi+i\left(J_{m+1}(\beta)-J_{m-1}(\beta)\right) \sin \phi \\
\left(J_{m+1}(\beta)+J_{m-1}(\beta)\right) \sin \phi-i\left(J_{m+1}(\beta)-J_{m-1}(\beta)\right) \cos \phi \\
~~~~~~~~~~~~~~~~~~~~~~~~~~~~~0
\end{array}\right] d \theta
\label{azi}
\end{aligned}
\end{multline}

\twocolumngrid

where, $\beta=k \rho sin \theta$, is the argument of Bessel functions, $\theta_{\max }=\sin ^{-1}(\mathrm{NA} / n)$, which is the maximum angle related to the NA of the objective, $n$ is the refractive index of the medium, $E^{o}$ is the output electric field, $ A $  is the constant related to the amplitude of the electric field and $f_{\omega}(\theta)$ is the apodization function that appears when an aplanatic lens tightly focuses the beam. $J_{m} $ is the Bessel function of $m^{th}$ order of the first kind, $\theta$ and $\phi $ denote the polar angle with respect to the z-axis and the azimuthal angle with respect to the x-axis in the cylindrical (or spherical) coordinate system, respectively. Subscripts (R) and (A) of the above equations \ref{rad} and \ref{azi} represent the radially and azimuthally polarized vector beams, respectively, and $\exp (i m \phi) $ represents the helical phase, where $m$ is the topological charge that can be any integer and $m \hbar$ is the OAM per photon. Just by putting $m = 0$ in the above equations \ref{rad} and \ref{azi}, we obtain the expressions of the first-order radially and azimuthally polarized vector beams. The tightly focused first-order ($m = 0$) radially polarized vector beam contains all three components of electric fields ($\mathbf{E}_{x}$, $\mathbf{E}_{y}$ and $\mathbf{E}_{z}$), with the transverse magnetic field components ($\mathbf{H}_{x}$ and $\mathbf{H}_{y}$), $\mathbf{H}_{z}$ being 0. However, an azimuthally polarized vector beam contains all three components of the magnetic field ($\mathbf{H}_{x}$, $\mathbf{H}_{y}$ and  $\mathbf{H}_{z}$), with the transverse electric field components ($\mathbf{E}_{x}$ and  $\mathbf{E}_{y}$), $\mathbf{E}_{z}$ being 0. The magnetic fields of radially polarized light are azimuthal in nature, whereas they are radial in nature for azimuthally polarized light. 

We now determine analytical expressions for the total OAM densities for tightly focused first-order ($m = 0$) radially and azimuthally polarized vector beams. For this, we first note that the linear momentum density $\mathbf{P}$ for a monochromatic electromagnetic field in SI unit is related to the time-averaged Poynting vector through $\mathbf{P}=\operatorname{Re}\left(\mathbf{E}^* \times \mathbf{H}\right) /\left(2 c^2\right)$ (or $\mathbf{P}=\operatorname{Re}\left(\mathbf{E} \times \mathbf{H}^*\right) /\left(2 c^2\right)$ ) \cite{emile2018energy,o2002intrinsic,xu2019azimuthal}. For the vector fields, we can decompose $\mathbf{P}$ into $\mathbf{P}^{o}$, which is the canonical or orbital momentum (CM) derived from the local phase gradient (wave vector) in the field, and a $\mathbf{P}^{s}$ (BSM) which arises as a consequence of spin inhomogeneity \cite{xu2019azimuthal, bliokh2014extraordinary}. Therefore, we can write $\mathbf{P}=\mathbf{P}^{o}+\mathbf{P}^{s}$, where $\mathbf{P}^{o}=\frac{1}{4 \omega n^2} \operatorname{Im}\left[\varepsilon \mathbf{E}^* \cdot(\nabla) \mathbf{E}+\mu \mathbf{H}^* \cdot(\nabla) \mathbf{H}\right]$ and $\mathbf{P}^{s}=\frac{1}{8 \omega n^2} \nabla \times \operatorname{Im}\left(\varepsilon \mathbf{E}^* \times \mathbf{E}+\mu \mathbf{H}^* \times \mathbf{H}\right)$, with $\epsilon$ being the permittivity,  $\mu$ the permeability, $n$ the refractive index of the embedding medium. The canonical $P^{o}$ (orbital) and spin $P^{s}$ parts of the momentum density generate the total OAM density of the electromagnetic field\cite{belinfante1940current,bliokh2014extraordinary,soper2008classical}. The total OAM density is $\bold{L=r \times P}$, where $r$ is the lateral position of the corresponding beam propagation axis \cite{o2002intrinsic}. Hence, the density of the linear momentum $\mathbf{P}$ and total OAM $\mathbf{L}$ for the radially (azimuthally) polarized vector beams on tight focusing in optical tweezers may be written as

\begin{equation}
\begin{aligned}
& P_x=\operatorname{Re}\left\{C\left(I_{12} I_{10}^{*}\right) \cos \phi\right\} \\
& P_y=\operatorname{Re}\left\{C\left(I_{12} I_{10}^{*}\right) \sin \phi\right\} \\
& P_z=\operatorname{Re}\left\{C\left(I_{12} I_{11}^{*}\right)\right\}
\end{aligned}
\label{Poynting}
\end{equation}

\begin{equation}
\begin{aligned}
& L_x=y P_z-z P_y \\
& L_y=z P_x-x P_z \\
& L_z=x P_y-y P_x
\end{aligned}
\label{OAM}
\end{equation}

 Here, $C$ is the constant denoting $\mathbf{P}$, while $I_{10}$, $I_{11}$, and $I_{12}$ represent the diffraction integrals (see equation 5 in Ref. \cite{kumar2023probing}) \cite{Novotny2012,roy2013controlled,kumar2021study,kumar2022probing,kumar2023probing}. Generally, Laguerre–Gaussian (LG$_{pl}$) beams \cite{yao2011orbital}, high-order Bessel beams \cite{durnin1987diffraction}, and Mathieu beams \cite{gutierrez2000alternative} inherently possess OAM. The mechanical action of any non-zero values of OAM ($m \neq 0$) causes particles to exhibit orbital motion around the beam propagation axis, as demonstrated in previous works\cite{he1995direct,garces2003observation, zhao2007spin}. However, in the case of beams with zero SAM ($\sigma=0$) and intrinsic OAM, such as first-order ($m=0$) radially and azimuthally polarized vector beams, particles still orbit around the beam center due to the spin-orbit interaction (SOI) effect. Consequently, the total OAM induces the rotation of birefringent particles around the beam propagation ($z$) axis. We will now proceed to numerically simulate our experimental system and evaluate the intensity distribution and OAM characteristics.

\begin{center}
\begin{figure}
\includegraphics[width=0.5\textwidth]{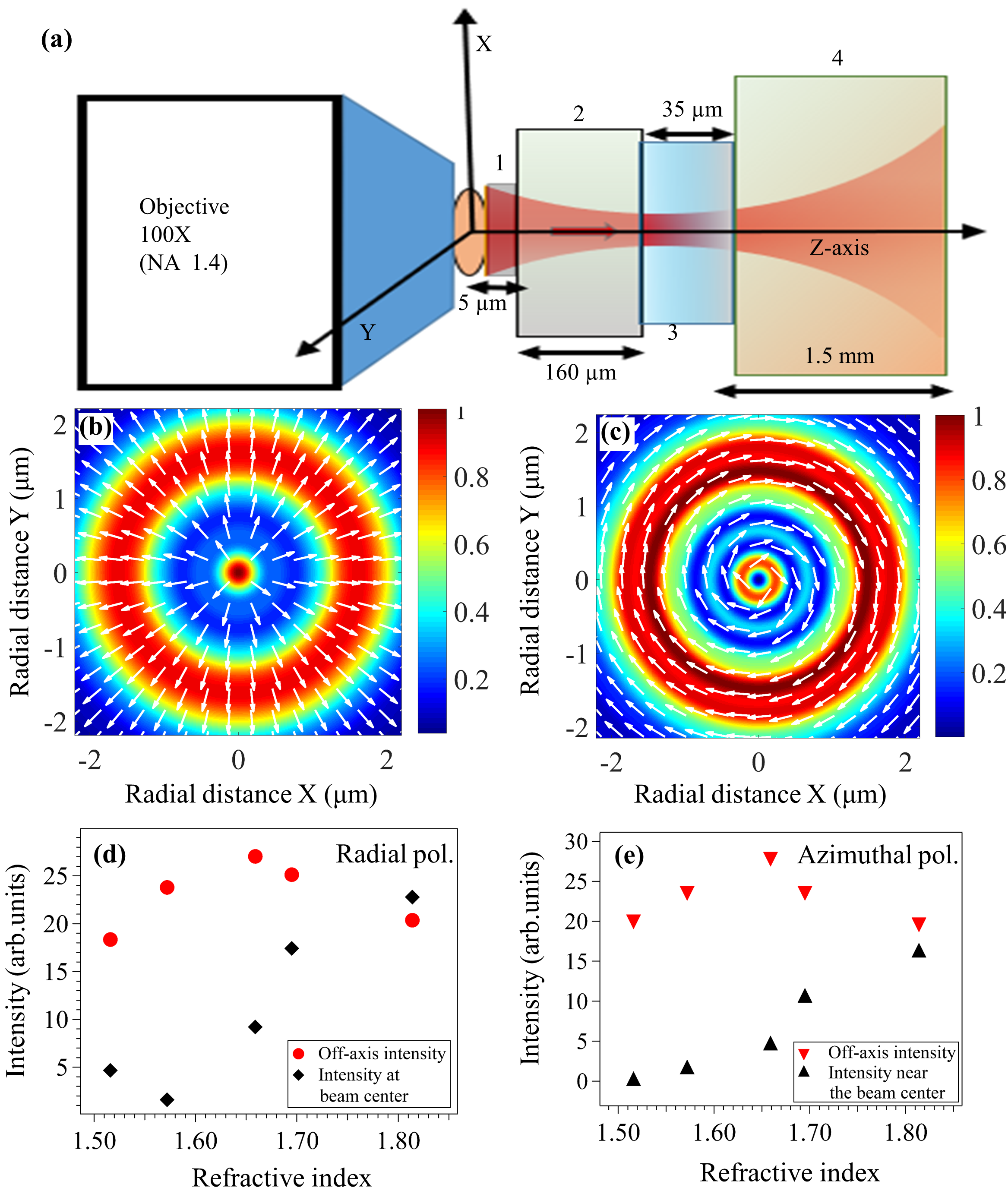}
\caption{(a) Schematic diagrams of the stratified medium used in our numerical analysis and in the optical tweezers system. Numerical simulation of the intensity distribution at z = 2 $\mu m$ away the focus for an RI 1.814 of the high NA objective (trap focus) lens for an input (b) radially polarized (c) azimuthally polarized vector beam. (d) and (e) Comparison of intensities at the beam center (or near the beam center) denoted by solid black squares and triangles, and off-axis annular ring represented by solid red circles and triangles as a function of RI at 2 $\mu m$ away from the focus for an input radially and azimuthally polarized vector beams, respectively.}
\label{Intensity}
\end{figure}
\end{center}

\section{ NUMERICAL SIMULATIONS}

In our experimental system, the output from a high NA objective lenses in optical tweezers setup is passed through a stratified medium. The laser beam of wavelength 671 nm is incident on the 100X oil immersion objective of NA 1.4 followed by (a) an oil layer of thickness around 5 $\mu m$ and refractive index (RI) 1.516, (b) a 160 $\mu m$ thick coverslip having refractive index varying between 1.516-1.814 (note that the case where the $RI = 1.516$ is henceforth referred to as the ``matched condition,” which is typically employed in optical tweezers to minimize spherical aberration effects in the focused beam spot, whereas the other values are referred to as a `mismatched' condition) (c) a sample chamber of an aqueous solution of birefringent RM257 particles in a water medium having a refractive index of 1.33 with a depth of 35 $\mu m$, and finally (d) a glass slide of refractive index 1.516 whose thickness we consider to be semi-infinite ( 1500 $\mu m$) [see Fig.~\ref{Intensity} (a)]. In the simulation, the origin of coordinates is taken inside the sample chamber at an axial distance of 5 $\mu m$ from the interface between the sample and the coverslip. Thus, the objective-oil interface is at -170 $\mu m$, the oil-cover slip interface is at -165 $\mu m$, and the cover slip-sample chamber interface is at -5 $\mu m$, and the sample chamber-glass slide interface is at +30 $\mu m$. 

\subsection{Numerical simulations of intensity distribution}
The schematic in Fig.~\ref{Intensity}(a) provides a cartoon representation of our system. We conduct the numerical simulations to investigate the tight focusing of the input radial/azimuthal beam by a high NA objective lens into a stratified medium under both `matched' and `mismatched' conditions (though the experimental observations are for the mismatched case). Consistent with the theoretical predictions, we observe that the electric field intensity at the center of a radially polarized beam is a result of $I_{\text{long}}$ ($I_{10}$), while the off-axis intensity arises from $I_{\text{rad}}$ ($I_{11}$), as illustrated in Figs.~\ref{Intensity}(b) and (d). On the contrary, for an azimuthally polarized beam, as depicted in Figs.~\ref{Intensity}(c) and (e), the electric field intensity near the beam center and the off-axis annular ring are formed due to $I_{\text{azi}}$ ($I_{12}$). In particular, at the center of the beam, the electric field intensity is found to be zero due to the absence of the longitudinal component ($E_{z}$) of the electric field. It is evident from Figs.~\ref{Intensity} (d) and (e) that, for input radial and azimuthal polarizations, the total intensity (resulting from both electric and magnetic fields) at the beam center and near the beam center at an axial distance of 2 $\mu$m from the focus can be increased by increasing the refractive index contrast of the stratified medium. For RI 1.814,  the intensity at the beam center is approximately 10\% higher than that in the annular ring for radial polarization, allowing particles to be trapped in both regions. 

\begin{center}
\begin{figure}
\includegraphics[width=0.5\textwidth]{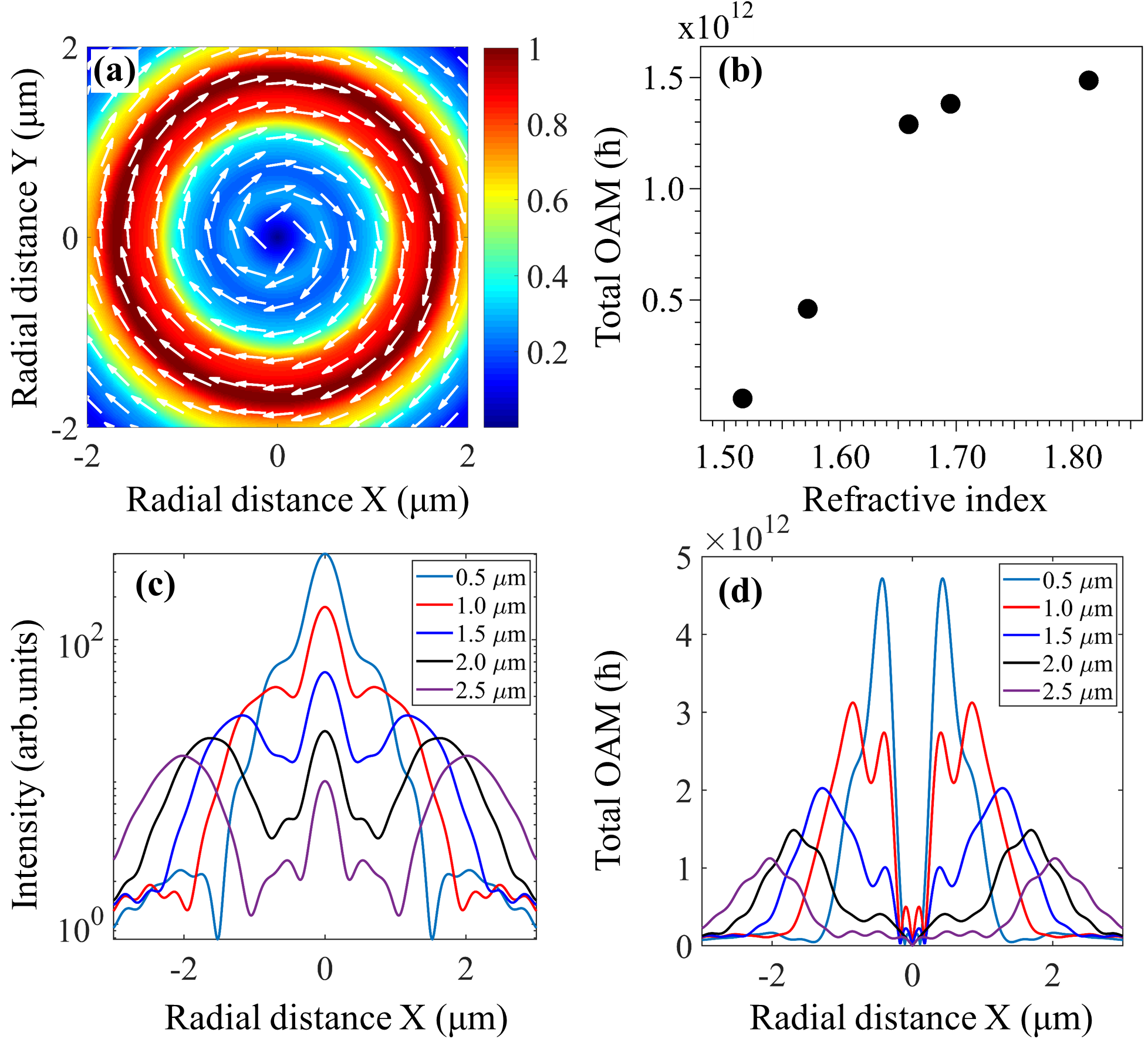}
\caption{ (a) Quiver plot of the origin-dependent OAM at $2~ \mu m$ away from the focus of a radially (azimuthally) polarized vector beam for mismatched RI (1.814). Note that the magnitude of the OAM for both beams is the same. (b) The maximum value of OAM is a function of RI. It is evident that the OAM is at its maximum at an RI of 1.814, which is consistent with our experimental conditions. Line plot of the radial intensity and total OAM distribution of a radially polarized beam at different $z$ planes for mismatched RI (1.814), shown in (c) and (d), respectively.}
\label{theo_OAM}
\end{figure}
\end{center}

\subsection{Numerical simulations of OAM distribution}
The results of our simulations correspond to origin-dependent total OAM are shown in Fig. ~\ref{theo_OAM} (a)-(d). For OAM (Fig. ~\ref{theo_OAM} (a), (b) and (d)) and the radial distribution of intensity (Fig. ~\ref{theo_OAM} (c)), we show results for the mismatched RI - since the value of OAM is highest for an RI of 1.814, which we show in Fig. ~\ref{theo_OAM} (b). Also, the spherically aberrated intensity profile that we obtain in this case allows overlap between significant intensity and large OAM that is useful to see effects on mesoscopic particles of a diameter of a few microns \cite{kumar2023probing, pal2020direct}. Note that we also perform the simulations not at the focal region of the trap, but at 2 $\mu m$ away from the focus - so as to obtain enough spatial extent of both intensity and OAM to obtain experimentally discernible effects. The corresponding intensity distributions as a function of axial distance from focus for the mismatched RI 1.814 are shown in Fig. ~\ref{theo_OAM} (c) shows that as we move away from the focus, off-axis intensity lobes are formed. The quiver plot in Fig.~\ref{theo_OAM}(a) illustrates that the total OAM for the mismatched RI case follows a clockwise direction and is distributed in an annular ring around the focus. We envisage that particles trapped in this ring would exhibit rotational motion around the beam center. In this note, Fig.~\ref{theo_OAM}(b) shows that the maximum value of OAM distributions at 2 $\mu m$ away from the focus is a function of RI. It is clear that the OAM is at its maximum at an RI of 1.814, which is what we use in our experiments. Fig.~\ref{theo_OAM}(d) demonstrates the radial distributions of total OAM as a function of axial distance from focus for the mismatched RI 1.814.  From this, it is clear that the OAM peaks overlap with off-axis intensity peaks (see Fig.~\ref{theo_OAM} (c)), suggesting that birefringent RM257 particles trapped on the off-axis would feel its effects the most.

 \subsection{Ellipticity Generation} 
 \begin{figure}[!h]
	\centering
	\includegraphics[scale=0.45]{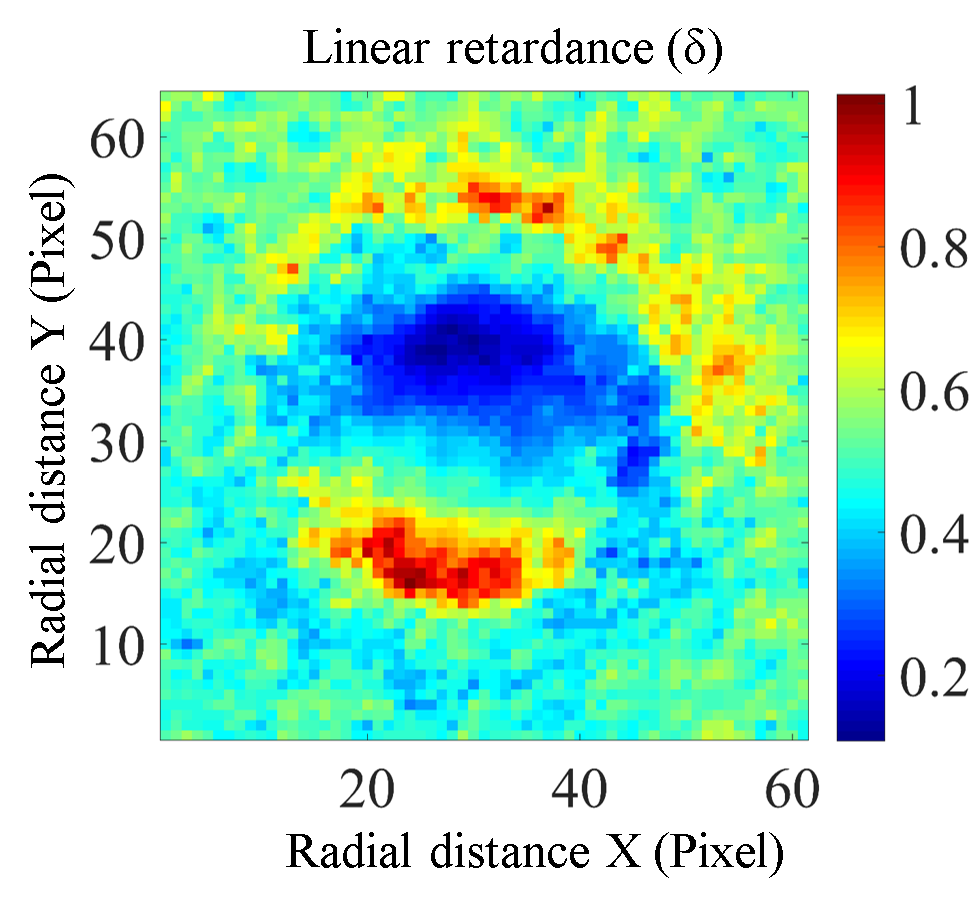}
	\caption{ The linear retardance of RM257 birefringent particle} 
	\label{retardance}
\end{figure}  

\begin{figure}[h]
	\centering
	\includegraphics[width=0.5\textwidth]{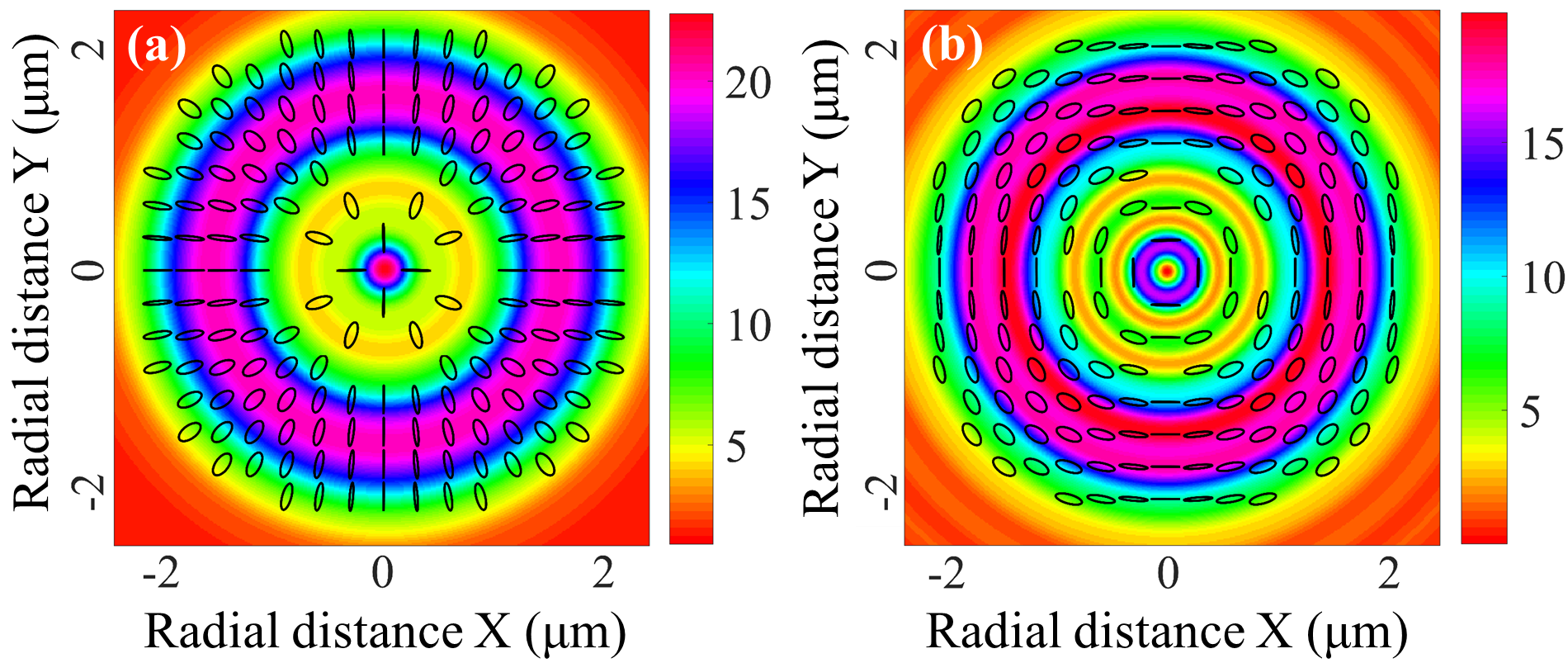}
	\caption{ (a) and (b) are the polarization ellipse plots of the emerging radially and 
            azimuthally polarized field from the RM257 birefringent particles, respectively.} 
	\label{pol ellipse}
\end{figure}

It is important to note that the first-order radially and azimuthally polarized vector beams inherently do not possess any SAM and OAM. However, when interacting with birefringent RM257 particles within a tightly focused beam in the stratified medium, ellipticity is induced in the output beam. Figure ~\ref{retardance} illustrates the spatial variation of the linear retardance ($\delta$) of the RM257 particles. The standard Jones calculations of focused radially and azimuthally polarized vector beams on the birefringent Rm257 particle are given as
\begin{equation}
\begin{aligned}
& \left(\begin{array}{ll}
1 & 0 \\
0 & e^{i \delta}
\end{array}\right)\left(\begin{array}{ll}
-i I_{11}\cos \phi \\
-i I_{11}\sin \phi
\end{array}\right)=\left(\begin{array}{l}
-i I_{11} \cos \phi \\
-i I_{11} e^{i \delta} \sin \phi
\end{array}\right)_{Rad} \\
\end{aligned}
\label{rad ellip}
\end{equation}

\begin{equation}
\begin{aligned}
& \left(\begin{array}{ll}
1 & 0 \\
0 & e^{i \delta}
\end{array}\right)\left(\begin{array}{cc}
i I_{12} \sin \phi \\
-i I_{12}\cos \phi
\end{array}\right)=\left(\begin{array}{l}
+i I_{12} \sin \phi \\
-i I_{12} e^{i \delta} \cos \phi,
\end{array}\right)_{Azi}
\end{aligned}
\label{azi ellip}
\end{equation}
where $\delta$ is the linear retardance of the RM257 particles, Eqs.~\ref{rad ellip} and ~\ref{azi ellip} are the resultant Jones vectors of the focused output field of radially and azimuthally polarized vector beams, respectively. We take the maximum value of the linear retardance ($\delta$) of the RM257 particles and calculate the Stokes vectors using Eqs.~\ref{rad ellip} and ~\ref{azi ellip}. Using the Stokes vector analysis, we plotted the polarization ellipse of the emerging radially and azimuthally polarized field from the particles shown in Figs.~\ref{pol ellipse}(a) and (b), respectively. This observation leads us to the conclusion that the contribution of BSM ($P^{s}$) is also crucial in the calculation of total OAM. Consequently, it is clear that birefringent particles would be experiencing a higher total OAM, thus enabling their rotation around the beam axis.


\section{Conclusion}
In conclusion, we use the SOI of light generated due to the tight focusing of structured vector beams in optical tweezers and its modification in the presence of interactions with birefringent micro-particles to engineer orbital rotation in the particles. Thus, we tightly focus radially and azimuthally polarized vector beams that do not carry any SAM and intrinsic OAM into a RI-stratified medium, and observe the effect of total OAM on birefringent particles orbiting around the beam propagation axis. We see clear signatures of origin-dependent OAM generated for both input polarizations. These effects are also facilitated by the spherical aberrated intensity profile generated by our RI stratified medium. Most importantly, the interaction of the focused light with the birefringent particles results in the generation of a finite BSM that actually enables the orbital rotation, which - importantly - is not observed in non-birefringent particles. Consequently, our work demonstrates how light-matter interactions may modify the effects of the SOI of light in optical tweezers to generate interesting dynamics in trapped particles.  Importantly, these effects arise from the SOI generated by tight focusing alone, without the need to structure complex beam profiles using advanced algorithms involving adaptive optics. In the future, we plan to observe the effects on SOI due to tight focusing, stratification, and light-matter interactions employing more complex structured beams. We even intend to extend our work on ENZ (Epsilon Near Zero) materials \cite{eismann2022enhanced}, to devise interesting routes for generating complex particle trajectories in optical tweezers.

\section{Acknowledgments:}
The authors acknowledge the SERB, Department of Science and Technology, Government of India (Project No. EMR/2017/001456). RNK acknowledges IISER Kolkata for providing research fellowship.



\providecommand{\noopsort}[1]{}\providecommand{\singleletter}[1]{#1}

\end{document}